\documentclass[graybox]{svmult}

\usepackage{natbib}

\usepackage{helvet}         
\usepackage{courier}        
\usepackage{type1cm}        
\usepackage{makeidx}         
\usepackage{graphicx}        
\usepackage{multicol}        
\usepackage[bottom]{footmisc}

\makeindex             

\begin{document}

\title*{Challenges for LSST scale data sets}
\author{Michael J. Way}
\institute{Michael J. Way \at NASA Goddard Institute for Space Studies,
2880 Broadway, New York, NY, USA, \email{michael.j.way@nasa.gov}}
%
%
\maketitle

\abstract*{The Large Synoptic Survey Telescope (LSST) simulator being built by
Andy Connolly and collaborators is an impressive undertaking and should
make working with LSST in the beginning stages far more easy than it
was initially with the Sloan Digital Sky Survey (SDSS). However, I would like
to focus on an equally important problem that has not yet been discussed here,
but in the coming years the community will need to address -- can we deal
with the flood of data from LSST and will we need to rethink the way we work?}

\abstract{The Large Synoptic Survey Telescope (LSST) simulator being built by
Andy Connolly and collaborators is an impressive undertaking and should
make working with LSST in the beginning stages far more easy than it
was initially with the Sloan Digital Sky Survey (SDSS). However, I would like
to focus on an equally important problem that has not yet been discussed here,
but in the coming years the community will need to address -- can we deal
with the flood of data from LSST and will we need to rethink the way we work?}

\section{Changing the way we work: From 2MASS to SDSS}
\label{sec:1}

Perhaps the best way to start things is to compare two large area sky
surveys implementing the ``standard way'' of distributing data in their own
time: The 1990s era Two Micron All Sky Survey \citep[2MASS;][]{Skrutskie06}
and the 2000s era Sloan Digital Sky Survey \citep[SDSS;][]{York2000}.

Initially if a researcher wanted to access the
2MASS\footnote{http://www.ipac.caltech.edu/2mass} survey data one
could obtain a 5 DVD set (double-sided) of the catalog data.  The data were 
bar-delimited ascii text which could be easily read by everything from
legacy scripting programs like awk to SQL databases like MySQL or
Postgres. The ascii source catalog was about 43GB in size if copied
from the DVDs to local disk. The full-fidelity Atlas Images ($\sim$10
terabytes in size and not available via DVD) were later accessible via
the 2MASS Image Services
website\footnote{http://irsa.ipac.caltech.edu/applications/2MASS/IM}.

In essence, the average astronomer had to change almost nothing about the way
that they or their Ph.D. advisor had worked over the previous 30 years.
For example, instead of ordering 9-track (1/2 inch = 12.7mm),
exabyte (8mm), or DDS/DAT (4mm) tapes from the observatory (or bringing
them along after an observing run) one simply ordered the 2MASS DVDs.
This was possible due to increases in computer cpu speed and
memory capacity combined with modest input/output (IO) improvements
over the previous 3-4 decades.

All of this changed with the SDSS. It may have been possible to distribute
DVD copies of the SDSS in a similar way to that of the 2MASS, but the scale
had moved from gigabytes to terabytes.  Having a few terabytes on a local
computer in the early 2000s was not common, so the SDSS 
took a different route. Working with top-notch computer scientists such
as Jim Gray of Microsoft they decided that much of the SDSS should be accessible
via SQL query.  There was certainly some anxiety amongst much of the community
when they realized that their mode of obtaining data from the SDSS was going
to be markedly different than in the past. Hence, it took the community some
time to learn this new way of working,
and certainly some early publications with SDSS data
not published by the SDSS team were problematic because, for example, they
did not realize that they could decide the quality of the photometry at a
fine level, unlike that of the 2MASS which was relatively straightforward.

The SDSS is probably the most similar data set today compared with what
the LSST will look like and how we will interact with it.
Currently most users of SDSS use the casjobs\footnote{http://casjobs.sdsss.org}
interface to obtain their data. The back end is a SQL database tied to a front
end presented to the user as a web interface where SQL queries are entered.
The database comes with a
Schema Browser\footnote{http://casjobs.sdss.org/dr7/en/help/browser/browser.asp}
that allows one to explore items of possible interest.
There are also a host of on-line tutorials that one can go through to understand
how make the queries, and many authors also publish their SDSS queries
in the appendices of their publications. However, not enough authors do
the latter in my opinion, and hence it is often impossible to replicate the data
that people are using if the original author does not publish or cannot
recall their original query.

In the ten years since the creation of the SDSS disk data storage density
has continued its inexorable rise (see Sect.~\ref{sec:2}) and today one could
in fact host the entire SDSS database relatively easily and cheaply on a
modestly sized desktop computer (e.g. an off-the-shelf workstation with
4 $\times$ 2TB disks would do the trick). Again, one could (in theory)
dump the entire casjobs catalog to a giant ascii file akin to that of
the 2MASS and use awk or your favorite fortran program on it. One would
need a system that can use 64bit addressing, but that is fairly standard
today (2011). Of course the IO will make things slow ($\sim$ 4 hours
to read a 1TB disk sequentially), but nonetheless it is in theory quite
possible.  The question then arises, will one be able to work with LSST
in the same manner as the SDSS given the inexorable rise of faster CPUs,
Memory, and IO?

\section{Changing the way we work: From SDSS to LSST}
\label{sec:2}

As we consider the move from SDSS sized data sets to LSST
the questions that people like us might ask at this stage are
fairly straightforward:

\begin{enumerate}
\item{Will one be able to have a copy of the LSST data-set on a local desktop?
This will allow researchers to continue their pre-SDSS era methods of
data interrogation. This is what we like to call the {\it 2MASS mode.}}

\item{Will one still utilize a casjobs type web-query interface to obtain
LSST data of interest and then use legacy home-grown tools to work
on the data? This is what we call the {\it SDSS mode.}}

\item{Must one completely change the way one works with LSST scale data sets?
Will ``data locality" be required -- will one have to do all of the operations
to obtain a project's scientific goals on the database directly?
This may be called the {\it LSST mode}.}
\end{enumerate}

To attempt to answer these questions we have to look at several other
factors discussed in the following subsections.

\subsection{LSST Database Size and possible architecture}
\label{subsec:2.1}

We heard from Andy Connolly and Kirk Borne at this conference that the
LSST query database will be of order 10 petabytes (PB) in size,
while there will be around 60PB of images available after
10 years of operation. It turns out that query scales almost linearly
with the size of the database. Given historical trends in CPU, memory,
storage and IO this means one should be able to derive
catalogs and do joins on tables in a future LSST database as we do
today with SDSS casjobs. However, there are caveats related to IO that will
be discussed later.

While query scales linearly with database size, the same cannot be
said of the kinds of operations that scientists would prefer
to do on the data. For example, classfication, clustering, density
estimation are all normally O(N$^{2}$) or worse. However, earlier
today Alex Gray showed us that his group has managed to make
a host of algorithms O(N) that are normally considered to be O(N$^{2}$).

Regardless, this points to some interesting issues. Assuming petabyte database
sizes, the needs to do operations that are O(N$^{2}$), and the
need to look at a large fraction of the stored data (that will not fit in RAM)
how are researchers going to do these things on the LSST database
of tomorrow? Let's touch on the possible need for ``data locality''.
Normally one should only consider moving the data from the source of the data
if one needs more than 100,000 CPU cycles per byte of data \citep{BGS2006}.
The kinds of applications this brings to mind are Seti@HOME or cryptography.
Thankfully most science applications are more information intensive with
CPU to IO ratios less than 10,000 to 1. This means that we may have two reasons
to consider the possibility that we will not actually download
the data to our local machine/data-center: The size of the database
is too large (petabyte scale) and we have CPU to IO ration of less than 100,000
to 1.  We will address these issues in some detail in the next section,
but for now let's assume we will need to do some calculations at the site
of the database itself.

The LSST has teamed up with several industry partners to develop
a new database to host the LSST called
SciDB\footnote{http://www.scidb.org}.
The current plan is to host this database in several
different geographic locations (obviously to avoid a single point of
failure and to handle the anticipated load), but they also currently
plan to have an R interface for ``expert users'' As I mentioned
during my talk, I think this is an excellent idea, but I hope the
designers will consider adding additional languages such as Python
which currently has wrappers to support a host of useful tools
commonly used by the current generation of younger 
astronomers\footnote{For example, numpy, scipy, Rpy (R interface),
mlabwrap (MatLab), etc.}

\subsection{Moving the data around -- can I have a local copy and make
use of it?}
\label{subsec:2.2}

Will one be able to download and store the LSST database to a desktop
computer in 10 years time? If one wishes to download 1 petabyte over
a {\it dedicated} 1 gigabit/second line (in common use today)
it will take $\sim$100 days.  In 9 years let's assume everyone will
have 10 gigabit/second connections (the growth in desktop network speed
has not grown at the same accelerated rate of storage or CPU)
so that means it will only take about 10 days. That doesn't sound unreasonable.
Now one has to ask, how much will it cost to own 1 petabyte of storage?
One can look at historical trends documented in several places
on the internet to get some
idea\footnote{e.g. http://www.mkomo.com/cost-per-gigabyte}.
In Table \ref{tab:1} you can see what disks costs were 10 years ago,
today and by extrapolation in 10 years time.

\begin{table}
\caption{Storage cost historical trends$^a$}
\label{tab:1}       
\begin{tabular}{p{1cm}p{2cm}p{2cm}p{2cm}}
\hline\noalign{\smallskip}
Year & Cost/GB & Cost/TB & Cost/PB  \\
\noalign{\smallskip}\svhline\noalign{\smallskip}
2000 & \$19.00000 & \$19,000 & \$19 million\\
2010 & \$00.06000 & \$62 & \$62,000 \\
2020 & \$00.00002 & \$0.2 & \$200\\
\noalign{\smallskip}\hline\noalign{\smallskip}
\end{tabular}
\newline $^a$ Extrapolated from http://www.mkomo.com/cost-per-gigabyte
\end{table}

Looking at Table \ref{tab:1} one comes to the conclusion
that if one wants to keep a copy of the LSST data locally it should
not be a problem given the drop in price over time. After all, who
who would have imagined 15 years ago that they would be able to purchase
a 1 terabyte drive for their desktop computer for under \$100?

Unfortunately things are not this simple. To illustrate my point
I want to recall some more of Amdahl's rules of thumb for a balanced system.

\begin{enumerate}


\item{Bandwidth (BW): One bit of sequential IO/second per instruction/second}

\item{Memory: $\alpha$=1=MB/MIPS\footnote{Million Instructions Per Second}: one byte of memory per one instruction/sec}

\item{One IO operation per 50,000 instructions}

\end{enumerate}


Looking at Table \ref{tab:2} in the context of Amdahl's ROT perhaps the
biggest problem with high performance computer centers today and into the
near future is that they are CPU rich, but IO poor.
The cpus may spend a lot of time sitting idle while waiting for IO
to send them more to work on because not everything can be stored in RAM.
This problem is not going to go away, but
there are (thankfully) people aware of the issue who believe
that it is currently possible to keep power consumption low while
increasing sequential read IO throughput by an order of magnitude
using what are called Amdahl blades \citep{Szalay2010}.
Note that power consumption is becoming an issue for mid-level
data centers found at Universities and some goverment research labs.
Most of these don't have Google's electricity budget for powering them and
in fact many government data centers are even being shutdown
to save money\footnote{http://www.nytimes.com/2011/07/20/technology/us-to-close-800-computer-data-centers.html}.

\begin{table}
\caption{Conclusions from Amdahl's rules of thumb for a balanced system today}
\label{tab:2}       
\begin{tabular}{p{2cm}p{2cm}p{2cm}p{1cm}p{3cm}}
\hline\noalign{\smallskip}
Operations & RAM & Disk I/O & &  No. of Disks for that \\
per second &     & bytes/s  & & BW at 100MB/s/disk    \\
\noalign{\smallskip}\svhline\noalign{\smallskip}
Giga/10$^{9}$  & GB & 10$^{8}$  & $\rightarrow$ & 1          \\
Tera/10$^{12}$ & TB & 10$^{11}$ & $\rightarrow$ & 1000       \\
Peta/10$^{15}$ & PB & 10$^{14}$ & $\rightarrow$ & 1,000,000  \\
\noalign{\smallskip}\hline\noalign{\smallskip}
\end{tabular}
\end{table}

Table \ref{tab:2}
tells one a couple of other interesting things.\footnote{This table is a modified version of one given in a talk by Alex Szalay
that the author became aware of recently.}  First, for a
Peta-scale balanced system 100TB/s of IO bandwidth (last row of column
three = 10$^{14}$) would be required.
It will take approximately 1,000,000 disks to deliver this bandwidth {\it today}
assuming they are capable of 100MB/s/disk. Note that the rate of disk IO growth
has not been remarkable in the past 10 years \citep[see][]{Szalay2010}.

\section{Conclusions}
\label{sec:conclusions}

In the beginning of Section \ref{sec:2} I posed three questions
and I would like to pose some answers:

\begin{enumerate}
\item{Will one be able to have a copy of the LSST data-set on a local desktop?
Yes, I think the average researcher will be able to have a copy of the data
on their local system assuming disk storage density continues its
historical trend (note that there are a number of technical arguments against
this, as there are for continuing Moore's law into the
future \citep{Hadi2011}\footnote{http://www.nytimes.com/2011/08/01/science/01chips.html}.
However, even if one has a copy of the LSST it is unlikely one will be able
to make much use of it using traditional {\it 2MASS mode} tools given
the issues with sequential IO that were outlined above.}

\item{Will one still utilize a casjobs type web-query interface to obtain
LSST data of interest and then use legacy home-grown tools to work
on the data (The {\it SDSS mode.})? Yes, the LSST team seems interested in
continuing the use of a casjobs type interface with an SQL backend.
Whether a researcher will then be able to use their traditional home-grown
tools will depend on the data sizes they download as discussed above.}

\item{Must one completely change the way one works with LSST scale data sets?
I believe that many of the scientific goals will only be
achievable by running on the database locally as an ``expert user''.
This points to the need to have a multitude of robust data/computational
centers hosting the LSST data. Today the best place for these (in the United
States) would probably be at the national level supercomputing centers
such as PSC\footnote{Pittsburg Supercomputing Center},
NSCA\footnote{National Center for Supercomputer Applications in Illinois} or
NAS\footnote{NASA Advanced Supercomputing center at NASA/Ames in California}
to name a few in the USA.}

\end{enumerate}

\begin{acknowledgement}
Thanks to Andy Connolly for taking the time to discus his LSST simulator
with me prior to the conference and for encouraging me in my belief that
a commentary focused on computational challenges would be appropriate.
I would also like to thank the Astrophysics Department at Uppsala University
in Sweden for their gracious hospitality while part of this manuscript was
being completed.
\end{acknowledgement}

\end{document}